\documentclass[conference]{IEEEtran}
\IEEEoverridecommandlockouts

\usepackage{graphicx}
\usepackage{listings}
\usepackage{hyperref}
\usepackage{xcolor}
\usepackage{amsmath}
\usepackage{url}
\usepackage{tikz}
\usetikzlibrary{arrows.meta,positioning,shadows.blur,shapes.geometric,fit,backgrounds,calc}

\lstdefinelanguage{yaml}{
  keywords={true,false,null,y,n},
  keywordstyle=\color{blue!70!black}\bfseries,
  basicstyle=\ttfamily\footnotesize,
  sensitive=false,
  comment=[l]{\#},
  commentstyle=\color{gray!70!black}\itshape,
  stringstyle=\color{orange!60!black},
  showstringspaces=false,
  breaklines=true,
  frame=single,
  backgroundcolor=\color{gray!5},
}

\begin{document}

\title{GITER: A Git-Based Declarative Exchange Model Using Kubernetes-Style Custom Resources }

\author{
\IEEEauthorblockN{Christos Tranoris}
\IEEEauthorblockA{Electrical and Computing Engineering Dpt, University of Patras, Greece\\
Email: \href{mailto:christos.tranoris@upatras.gr}{tranoris@ece.upatras.gr}\\
ORCID: 0000-0002-3433-037X\\
Published: Nov 2025}
}

\maketitle

\begin{abstract}
This paper introduces a lightweight and auditable method for asynchronous information exchange between distributed entities using Git as the coordination medium. 
The proposed approach replaces traditional APIs and message brokers with a \textbf{Git-based communication model} built on the principles of Kubernetes Operators and Custom Resources (CRs). 
Each participating entity—designated as a \textit{Publisher} or \textit{Consumer}—interacts through a shared repository that serves as a single source of truth, where the \texttt{spec} field captures the desired state and the \texttt{status} field reflects the observed outcome. 
This pattern extends GitOps beyond infrastructure management to support cross-domain, inter-organizational, and air-gapped collaboration scenarios. 
By leveraging Git’s native features (versioning, commit signing, and access control) the model ensures transparency, traceability, and reproducibility while preserving loose coupling and autonomy between systems. 
The paper discusses architectural principles, implementation considerations, and comparisons with RESTful and broker-based integrations, highlighting both the advantages and trade-offs of adopting Git as a declarative communication substrate.
\end{abstract}

\section{Introduction}
Modern distributed systems often require secure, auditable, and asynchronous exchanges between components that may not share live, direct connectivity. 
This white paper presents a Git-based exchange model that enables two entities---a \textit{Publisher} and a \textit{Consumer}---to communicate declaratively through a shared Git repository, which serves as a synchronized state store that both sides continuously reconcile against. 

Rather than relying on APIs or message brokers, the two entities exchange information via a single Custom Resource (CR) file committed to the repository. 
This resource conforms to a \textit{Custom Resource Definition (CRD)} schema that distinguishes between the \texttt{spec} (desired state) written by the Publisher and the \texttt{status} (observed state) updated by the Consumer.

The proposed pattern is lightweight, transparent, and inherently secure, making it particularly suitable for loosely coupled integrations across organizational, system, or network boundaries where persistent connectivity is neither possible nor desirable. 
By combining GitOps principles with the Kubernetes Operator reconciliation model, this approach extends declarative automation to inter-entity communication while preserving full auditability and traceability through Git history.

\section{Background and Motivation}

GitOps has revolutionized the management of cloud-native systems by treating Git as the single source of truth for system state. 
In a traditional GitOps workflow, an operator or deployment controller (ArgoCD, FluxCD) observes the desired configuration stored in Git and applies it to the target system (e.g., a Kubernetes cluster). 
However, this model is fundamentally \textit{one-directional}: the user commits a change to Git to declare a desired state, but Git itself does not provide a native mechanism to return or synchronize the actual \textit{status} of the system back to the user. 
The reconciliation happens in the background, and users typically learn about the system state through external monitoring, dashboards, or CLI tools—not through the same Git channel that triggered the change.

In parallel, \textit{RESTful APIs} and \textit{message brokers} (such as ActiveMQ, Kafka, or NATS) are the most common means of machine-to-machine communication. 
While they provide immediate feedback and event-driven interactions, they introduce operational coupling, authentication complexity, and availability dependencies. 
In contrast, Git-based exchanges are inherently \textit{asynchronous}, \textit{auditable}, and \textit{declarative}, allowing participating entities to remain independent and even offline between interactions.

\smallskip
By merging these ideas, we obtain a pattern where two entities interact entirely through Git, exchanging structured declarative resources with the same clarity, auditability, and automation as GitOps deployments, yet now with bidirectional visibility of both desired and observed state. 
Instead of defining a new custom protocol or message format, this approach adopts the well-established \textit{Kubernetes Custom Resource (CR)} model, where each resource has a \texttt{spec} (desired state) and a \texttt{status} (observed state). 
The producer modifies only the \texttt{spec}, the consumer modifies only the \texttt{status}, and both sides reconcile the Git repository to ensure convergence. Here are some key motivations:

\begin{itemize}
  \item \textbf{Protocol reuse:} instead of defining a new ad-hoc message format, the model reuses the proven and widely adopted Kubernetes CRD structure.
  \item \textbf{Air-gapped and asynchronous operation:} no live APIs or message brokers are required; entities can act independently and reconcile periodically.
  \item \textbf{Complete traceability:} all exchanges are recorded in Git commits and diffs, ensuring full auditability.
  \item \textbf{Declarative interface:} clear separation between desired and observed state via \texttt{spec}/\texttt{status}.
  \item \textbf{Autonomous reconciliation loops:} both sides independently align their local state with the shared Git resource.
  \item \textbf{Security and governance:} built-in Git access control, commit signatures, and branch protection provide trust and accountability.
\end{itemize}

\subsection*{The Kubernetes Operator Pattern and Custom Resources}

Kubernetes extends its declarative management model beyond built-in objects (such as Pods, Deployments, or Services) through the \textit{Operator pattern}. 
An Operator encapsulates domain-specific operational knowledge in the form of a controller that continuously reconciles the desired configuration of a resource with the current state observed in the system. 
This reconciliation loop embodies the core control theory principle of feedback: it reads the declared state, compares it against reality, and performs actions until convergence is achieved.

At the foundation of the Operator model are \textit{Custom Resources (CRs)} and their corresponding \textit{Custom Resource Definitions (CRDs)}. 
A CRD extends the Kubernetes API with new resource types, allowing developers to represent any domain concept as a first-class object within the cluster. 
Each Custom Resource instance contains two key sections:
\begin{itemize}
  \item \texttt{spec} --- the \textit{desired state}, expressed declaratively by the user or an external system.
  \item \texttt{status} --- the \textit{observed state}, maintained and updated by the controller to reflect runtime conditions or outcomes.
\end{itemize}

The Operator continuously monitors Custom Resources, interpreting changes in the \texttt{spec} and adjusting the underlying system to make the observed \texttt{status} match the desired configuration. 
This feedback loop enables automation of complex life-cycle tasks (such as installation, scaling, healing, and upgrades) through declarative definitions rather than imperative scripts or manual interventions.

By leveraging this pattern outside the boundaries of a Kubernetes cluster, the same reconciliation principles can govern interactions between distributed entities. 
When combined with Git as the synchronization medium, the Operator model provides a familiar and robust framework for \textit{declarative, auditable, and autonomous coordination} between independent systems.

\section{Conceptual Model}
The system comprises:
\begin{itemize}
  \item \textbf{Producer}: The entity that initiates the request. Expresses the desired state by writing/updating only the \texttt{spec} sections
  \item \textbf{Consumer}: The entity that processes the request. Interprets \texttt{spec}, executes work, Writes, and updates only the  \texttt{status} section.
  \item \textbf{Git Repository}: shared source of truth, history and audit trail.
\end{itemize}

Both entities (Producer/Consumer) reconcile the Git repository periodically or continuously, much like Kubernetes controllers watching Custom Resources.

\subsection{Resource Structure}
\begin{lstlisting}[language=yaml,caption={Example Custom Resource YAML}]

apiVersion: exchange.gitops/v1alpha1
kind: TaskExchange
metadata:
  name: example-task
spec:
  action: "process-video"
  parameters:
    inputUrl: "https://example.com/video.mp4"
status:
  phase: "Pending"
  result: {}
\end{lstlisting}

The producer updates only \texttt{spec}; the consumer updates only \texttt{status}.

\begin{figure}
    \centering
    \includegraphics[width=1.0\linewidth]{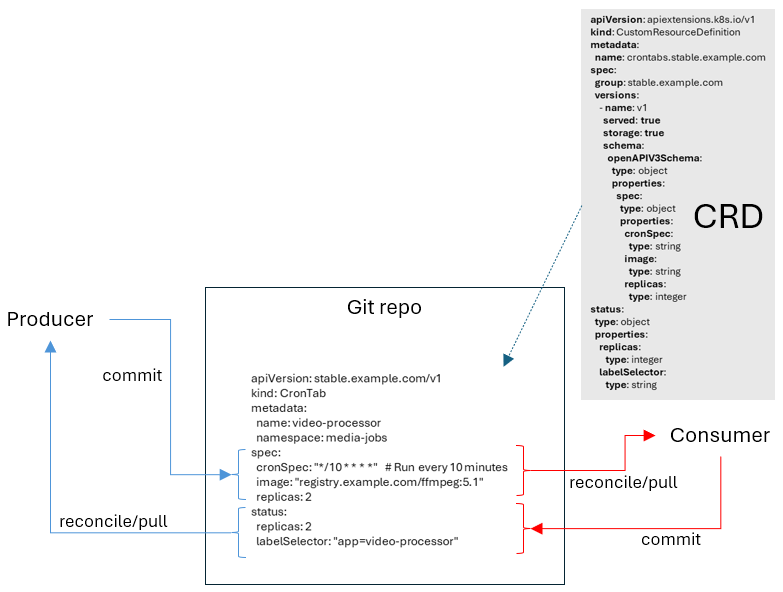}
    \caption{GITER pattern}
    \label{fig:placeholder}
\end{figure}

\subsection{Exchange Workflow}
Communication occurs asynchronously, through Git commits and push/pull operations, rather than live API calls. The workflow is as follows: 

\begin{enumerate}
  \item \textbf{Initialization:} Both entities have read/write access to a shared repository and agree on the CRD.
  \item \textbf{Producer commits CR:} Creates/updates a file (e.g., under \texttt{/resources/}); modifies only \texttt{spec}; pushes.
  \item \textbf{Consumer reconciliation loop:} Pulls; reads \texttt{spec}; processes; writes \texttt{status}; pushes.
  \item \textbf{Producer reconciliation loop:} Pulls; observes \texttt{status}; proceeds (e.g., archive/delete on completion).
  \item \textbf{Lifecycle termination:} Completed CRs are archived or removed.
\end{enumerate}

\section{Discussion}

\subsection{Benefits}

The proposed Git-based exchange model inherits several advantages from both GitOps and the Kubernetes Operator reconciliation paradigm. 
It provides a unified, declarative, and fully auditable mechanism for coordinating actions across distributed entities without requiring continuous connectivity or bespoke integration logic.

\begin{itemize}
  \item \textbf{Auditability:} Every exchange is captured as a Git commit, complete with versioning, timestamps, and optional cryptographic signatures. This ensures full accountability and non-repudiation of changes.
  \item \textbf{Asynchrony:} Communication occurs through the repository, allowing entities to operate independently and reconcile at their own pace without direct or persistent network connections.
  \item \textbf{Simplicity:} The model eliminates the need for message brokers, queues, or complex API endpoints, relying solely on the well-understood Git workflow of pull, commit, and push.
  \item \textbf{Traceability:} The entire lifecycle of an exchange: creation, processing, and completion is visible in the Git history, enabling precise auditing and rollback capabilities.
  \item \textbf{Security:} Authentication, access control, and commit signing are inherited from Git’s mature security mechanisms, while branch protection policies provide additional safeguards.
  \item \textbf{Reproducibility:} Because every state transition is stored in Git, the system’s evolution can be replayed, analyzed, or reconstructed at any point in time, ensuring deterministic outcomes.
\end{itemize}

\subsection{Potential Use Cases}

The proposed Git-based exchange model is applicable in a wide range of distributed automation and coordination scenarios where direct communication channels are restricted or undesirable. 
Its asynchronous, declarative, and auditable nature makes it particularly valuable in cross-domain or multi-organizational contexts that demand trust, autonomy, and verifiable state management.

\begin{itemize}
  \item \textbf{Inter-organizational automation without shared infrastructure:} 
  Two or more organizations can coordinate workflows or data exchanges through a shared Git repository, without exposing internal APIs or requiring VPNs or service meshes. 
  Each party maintains control of its own environment while achieving transparent and verifiable collaboration.

  \item \textbf{Federated orchestration and configuration propagation:} 
  In multi-domain or multi-cluster environments, configurations or service intents can be propagated declaratively across independent orchestration domains. 
  The Git repository acts as the federation hub, ensuring that each domain reconciles a consistent and traceable configuration.

  \item \textbf{Air-gapped and disconnected operations:} 
  Systems operating in restricted or offline environments—such as industrial networks, defense systems, or remote testbeds—can exchange state information through controlled Git synchronization (e.g., periodic pull/push via removable media or scheduled replication).

  \item \textbf{Distributed experiments and validation workflows:} 
  Research infrastructures and testbeds can share experiment descriptions, configurations, and results as versioned Custom Resources. 
  Each testbed acts as a Consumer that updates the \texttt{status} fields upon execution, while the experiment coordinator (Publisher) monitors progress entirely through Git.

  \item \textbf{Lightweight ``offline Kubernetes'' coordination:} 
  The model enables Operator-like behavior outside of Kubernetes clusters. 
  Entities can function as independent controllers, exchanging declarative state definitions through Git rather than relying on a live API server, ideal for edge, embedded, or hybrid deployments.
\end{itemize}

In all these cases, the Git repository provides a common, version-controlled coordination surface that preserves the declarative semantics of Kubernetes-style automation while operating securely and transparently across trust or connectivity boundaries.

\subsection{Implementation Considerations}

While conceptually straightforward, implementing the Git-based exchange model requires attention to practical aspects such as conflict management, repository organization, validation, and security. 
The following considerations summarize best practices and design recommendations for robust deployments.

\begin{itemize}
  \item \textbf{Conflict Resolution:} 
  To avoid merge conflicts and data corruption, it is essential to enforce a strict separation of concerns: the \texttt{spec} section is owned exclusively by the Producer (Publisher), while the \texttt{status} section is maintained solely by the Consumer. 
  Each side should update only its respective fields and perform merges using Git strategies that preserve this contract. 
  Optional pre-commit or CI checks can verify that updates do not cross these boundaries.

  \item \textbf{Branching Strategy:} 
  Different organizational or logical domains can be isolated through either per-exchange branches or dedicated subdirectories within the repository. 
  Branch-per-exchange workflows allow multiple interactions to proceed concurrently and independently, while directory-based namespacing simplifies aggregation of related resources. 
  Both approaches benefit from clear naming conventions and automated cleanup policies to manage completed exchanges.

  \item \textbf{Schema Enforcement:} 
  The Custom Resource Definition (CRD) schema serves as the formal contract between the Producer and Consumer. 
  Validation should occur automatically through continuous integration (CI) pipelines, ensuring that submitted CR files conform to the schema and contain only allowed fields. 
  This guarantees compatibility, prevents malformed input, and enables evolution of the schema through versioning.

  \item \textbf{Tooling:} 
  Lightweight reconciliation loops can be implemented using simple polling controllers written in Go, Python, or Java. 
  These controllers monitor the repository for new or updated CRs, apply domain-specific logic, and update the corresponding \texttt{status} fields. 
  Existing GitOps and Operator frameworks (such as \textit{FluxCD}, \textit{ArgoCD}, or the \textit{Operator SDK}) can be adapted to this pattern with minimal overhead.

  \item \textbf{Security:} 
  The repository should enforce strong authentication and authorization policies using Git’s built-in mechanisms or external identity providers. 
  Commit signing (via GPG or Sigstore) provides provenance and tamper evidence, while branch protection rules and code owners files can restrict modifications to critical areas. 
  When sensitive data must be exchanged, selective field encryption (e.g., using sealed secrets or SOPS) can maintain confidentiality without compromising transparency.
\end{itemize}

Properly addressing these implementation aspects ensures that the proposed model remains deterministic, verifiable, and secure, while maintaining the lightweight and transparent qualities that make it suitable for cross-domain collaboration.

\subsection{Comparison with Traditional Systems}

Traditional inter-system communication mechanisms, such as RESTful APIs and message brokers, offer mature and well-understood models for request/response or event-driven integration. 
However, these approaches inherently depend on continuous network availability, shared authentication infrastructures, and sometimes complex middleware. 
In contrast, the proposed GitOps-based Custom Resource (CR) exchange model replaces live interactions with asynchronous, version-controlled coordination over a Git repository.

\begin{table*}[!t]
\centering
\caption{Comparison of the GITER approach with Conventional Communication Paradigms}
\begin{tabular}{|p{3cm}|p{3.2cm}|p{3.2cm}|p{3.2cm}|}
\hline
\textbf{Feature} & \textbf{GITER} & \textbf{Message Queue} & \textbf{REST API} \\
\hline
Transport & Git & Broker & HTTP \\
\hline
Direction & Bidirectional & Unidirectional & Request/Response \\
\hline
Persistence & Versioned (in Git) & Typically Volatile & Stateless \\
\hline
Auditability & Built-in (Git history) & External Tools & External Tools \\
\hline
Offline Mode & Supported & No & No \\
\hline
Coupling & Loose & Medium & Tight \\
\hline
Security Model & Git-based Auth/Signatures & Broker ACLs & Token/OAuth \\
\hline
Schema Definition & CRD (YAML/JSON) & Custom Schema & OpenAPI \\
\hline
Best Suited For & Declarative, asynchronous coordination & Streaming and event propagation & Synchronous query or command \\
\hline
\end{tabular}
\end{table*}

\paragraph*{Advantages (Pros).}
The Git-based exchange model introduces several practical and architectural benefits:
\begin{itemize}
  \item \textbf{Decoupling and Asynchrony:} Entities operate independently without maintaining open network sessions or service endpoints. 
  This significantly simplifies integration across disconnected or partially trusted environments.
  \item \textbf{Inherent Auditability:} Every interaction is recorded in Git history, complete with timestamps, authorship, and optional digital signatures, providing an immutable audit trail.
  \item \textbf{Reproducibility and Version Control:} All exchanged states are versioned, enabling rollback, replay, and reproducible workflows.
  \item \textbf{Simplicity of Tooling:} Implementation relies solely on standard Git operations (commit, push, pull) without specialized middleware or message brokers.
  \item \textbf{Offline and Air-Gapped Operation:} The model remains functional even in environments with intermittent connectivity, as synchronization can occur periodically or through controlled replication.
  \item \textbf{Declarative Semantics:} The use of a CRD schema enforces clear contracts between entities, maintaining compatibility and simplifying validation.
\end{itemize}

\paragraph*{Limitations (Cons).}
Despite its advantages, the approach has certain trade-offs compared to conventional systems:
\begin{itemize}
  \item \textbf{Latency and Real-Time Constraints:} Since exchanges depend on Git synchronization cycles, near real-time communication is not feasible. 
  Message brokers and REST APIs remain preferable for time-critical applications.
  \item \textbf{Scalability:} Continuous Git operations (clones, fetches, merges) may introduce overhead for very large repositories or high-frequency updates.
  \item \textbf{Conflict Management:} Although separation of \texttt{spec} and \texttt{status} mitigates most conflicts, improper merges or concurrent commits can still occur and must be handled via automation or CI checks.
  \item \textbf{Operational Familiarity:} Organizations must adopt GitOps principles and repository-based workflows, which may differ from conventional API integration practices.
  \item \textbf{Event-Driven Limitations:} The model lacks the push-style event propagation that message brokers natively support, unless extended via Git webhooks or external triggers.
\end{itemize}

\section{Other considerations}
Here are some other generic consideration when adopting such approach:
\begin{itemize}
    \item \textbf{Multi-Party and Multi-File Exchanges}: Support multiple producers and consumers via namespaced directories and multiple CR instances; enforce role-based permissions delimiting who can edit \texttt{spec} vs.\ \texttt{status}.
    \item \textbf{Webhook and Event Triggers}: Use Git webhooks (e.g., GitHub/Gitea) to switch from polling to event-driven reconciliation for near real-time interactions.
    \item \textbf{Integration with OCI Registries}: Represent CRs as OCI artifacts to combine Git’s governance with scalability and mirroring of the registry.
    \item \textbf{Declarative Multi-Stage Pipelines}: Chain CRDs so one consumer’s \texttt{status} becomes another producer’s \texttt{spec}, forming declarative GitOps pipelines.
    \item \textbf{Security and Trust Enhancements}: Adopt Sigstore signing, OPA/Gatekeeper policies, and selected field encryption for zero-trust exchanges.
    \item \textbf{Performance and Scalability}: Investigate sparse checkouts, partial clones, and caching layers to support large and frequent updates.

\end{itemize}

\section{Conclusion}
We presented a GitOps-native pattern for declarative exchange between producer and consumer entities, modeled on Kubernetes-style reconciliation. The proposed GITER GitOps CR exchange model complements rather than replaces traditional communication mechanisms. 
It provides a highly auditable, asynchronous, and loosely coupled alternative well-suited for cross-domain coordination, disconnected environments, and scenarios demanding traceability over throughput. 
For interactive or high-frequency messaging, RESTful or broker-based solutions remain more appropriate, while Git-based exchanges excel in declarative, verifiable, and low-dependency integrations.

 Git provides the single source of truth with auditability and reproducibility; the \texttt{spec}/\texttt{status} split yields clean responsibilities. Challenges include event-driven reconciliation, multi-entity topologies, and trust frameworks, extending Git toward a federated orchestration substrate.

\bibliographystyle{IEEEtran}

\end{document}